\documentclass[a4paper]{jpconf}
\usepackage{graphicx}

\begin{document}
\title{Magnetic and electric properties 
in the distorted tetrahedral spin chain system Cu$_3$Mo$_2$O$_9$}

\author{T Hosaka$^1$, S Hachiuma$^1$, H Kuroe$^1$, T Sekine$^1$, M Hase$^2$, K Oka$^3$, T Ito$^3$, H Eisaki$^3$, 
M Fujisawa$^4$, S Okubo$^4$, H Ohta$^4$}
\address{$^1$Department of Physics, Sophia University, Tokyo 102-8554, Japan}
\address{$^2$National Institute for Materials Science (NIMS), Tsukuba, Ibaraki 305-0047, Japan}
\address{$^3$National Institute of Advanced Industrial Science and Technology (AIST), Tsukuba, Ibaraki 305-8568, Japan}
\address{$^4$Molecular Photoscience Research Center, Kobe University, Kobe, Hyogo 657-8568, Japan}
\ead{hosaka-t@sophia.ac.jp}

\begin{abstract}
 We study the multiferroic properties 
in the distorted tetrahedral quasi-one dimensional 
spin system Cu$_3$Mo$_2$O$_9$, in which 
the effects of the low dimensionality 
and the magnetic frustration 
are expected to appear simultaneously. 
We clarify that the antiferromagnetic order 
is formed together 
with ferroelectric properties at $T_{\rm N}=7.9$ K 
under zero magnetic field
and obtain the magnetic-field-temperature phase diagram 
by measuring dielectric constant 
and spontaneous electric polarization.
It is found that the antiferromagnetic phase possesses 
a spontaneous electric polarization
parallel to the $c$ axis  
when the magnetic field $H$ is applied parallel to the $a$ axis.  
On the other hand, 
there are three different ferroelectric phases 
in the antiferromagnetic phase 
for $H$ parallel to the $c$ axis.

\end{abstract}

\section{Introduction}

Recently, the discovery of the strong magnetoelectric effect
in TbMnO$_3$ \cite{Kimura2003} has rekindled significant interest in multiferroics displaying 
the interplay between ferromagnetic and ferroelectric properties.
After that, the multiferroism 
has been extensively studied \cite{Cheong2007} in transition metal oxides 
and a few microscopic mechanisms of this phenomenon have been proposed.
The inverse
Dzyaloshinskii-Moriya interaction 
and the inverse Kanamori-Goodenough interaction 
induce multiferroic properties in the spiral spin and 
collinear structures, respectively, where the magnetic
superlattices are formed.\cite{Katsura2005,Kenzelmann2005,Mostovoy2006}
The geometrical magnetic
frustration also plays an important role as the origin of
the nontrivial spin configuration which breaks the spatial inversion
symmetry.\cite{Arima2005} 
Recently we reported a possibility that Cu$_3$Mo$_2$O$_9$ shows
multiferroic behaviors {\it without} any magnetic superlattice formation.\cite{Kuroe2011-2}
In ref.\cite{Kuroe2011-2}, we focused on the dielectric properties and the electric polarization 
induced by an antiferromagnetic (AFM) spin order
when the magnetic field $H$ is applied along the $c$ axis.
In the present work, we report mainly the results for $H$ parallel to the $a$ axis.

Cu$_3$Mo$_2$O$_9$ has two distorted tetrahedral quasi-one dimensional
quantum spin systems made from $S = 1/2$ spins 
along the {\it b} axis in its orthorhombic unit cell (see Figs. 1(a) and (b)).\cite{Hamasaki2008,Hamasaki2010} 
This compound has magnetic
frustrations due to the tetrahedral spin alignment
and the quasi-one dimensionality simultaneously. This
compound undergoes an AFM phase transition at $T_{\rm N}$ = 7.9 K at zero magnetic field and shows a weak ferromagnetic order due to the spin canting at low temperatures.\cite{Hamasaki2008,Hamasaki2010} 
The inelastic neutron scattering study shows the hybridization effects due to the $J_1$ and $J_2$ superexchange interactions between two
elemental magnetic excitations, i.e., that of the quasi-one
dimensional AFM spin system made from the $J_4$
(= 4.0 meV) superexchange interactions and that of the isolated
AFM spin dimers made from the $J_3$ (= 5.8 meV)
ones.\cite{Kuroe2010,Kuroe2011}

\begin{figure}[h]
\includegraphics[width=22pc]{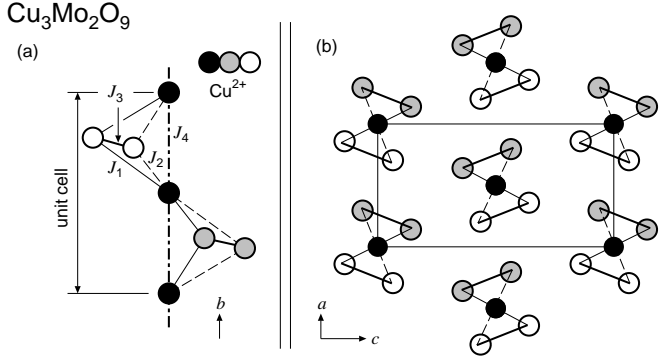}
\begin{minipage}[b]{16pc}\caption{\label{spin}Schematics of the distorted tetrahedral chain in
Cu$_3$Mo$_2$O$_9$ along the {\it b} axis (a) and in the {\it ac} plane (b). The
circles indicate the $S$ = 1/2 Cu$^{2+}$ ions and the symbols distinguish
their coordinates along the {\it b} axis from others. O$^{2-}$ and
Mo$^{4+}$ ions are omitted. The dashed, solid, bold and dot-dashed
lines distinguish the superexchange interactions $J_1-J_4$ between
Cu$^{2+}$ ions. The solid rectangle in (b) denotes the unit cell, which
contains two tetrahedral chains.}
\end{minipage}
\end{figure}

\section{Experiment}

We measured the temperature $T$ and magnetic-field $H$ dependences of the
dielectric constant $\epsilon_a$ and $\epsilon_c$ in Cu$_3$Mo$_2$O$_9$
when $H$ is applied along the $a$ and $c$ axes.
We prepared the plate-like single crystals of Cu$_3$Mo$_2$O$_9$ 
of which cross sections and thicknesses are typically about 60 mm$^2$ and
0.4 mm, respectively. 
To form a capacitor, the faces were
coated with gold and attached using gold wires. The capacitance,
of which the typical value was on the order
of 10 pF, was measured using the impedance analyzer
(Yokogawa-Hewlett-Packard 4192A).
$\epsilon_\alpha$ ($\alpha=a$ or $c$) was obtained
from the capacitance at 100 kHz with a peak
voltage of 1 V. 
The magnetic field was applied using a superconducting magnet (Oxford
Instruments, Teslatron S14/16), of which the maximum
magnetic field was 16 T.

\begin{figure}[b]
\includegraphics[width=38pc]{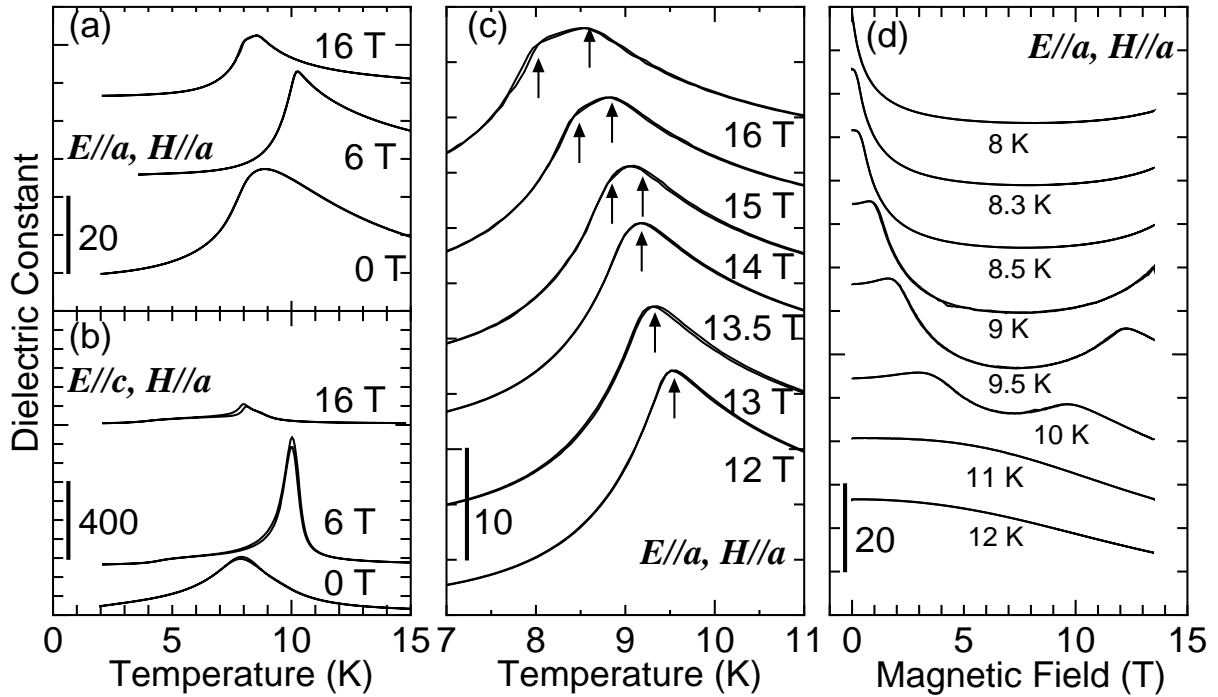}\hspace{1pc}%
\begin{minipage}{38pc}
\caption{\label{dielectric}
Typical temperature dependences of 
(a) the dielectric constant $\epsilon_a$, 
(c) its expansion between 12 and 16 T and 
(b) $\epsilon_c$ under fixed magnetic fields along the $a$ axis. 
For the visibility, the data were shifted. 
Arrows in (c) denote one and two peaks of dielectric constants.
The magnetic-field dependence of
the dielectric constant $\epsilon_a$ from 8 to 12 K is shown in (d).  
}
\end{minipage}
\end{figure}
\begin{figure}[t]
\includegraphics[width=38 pc]{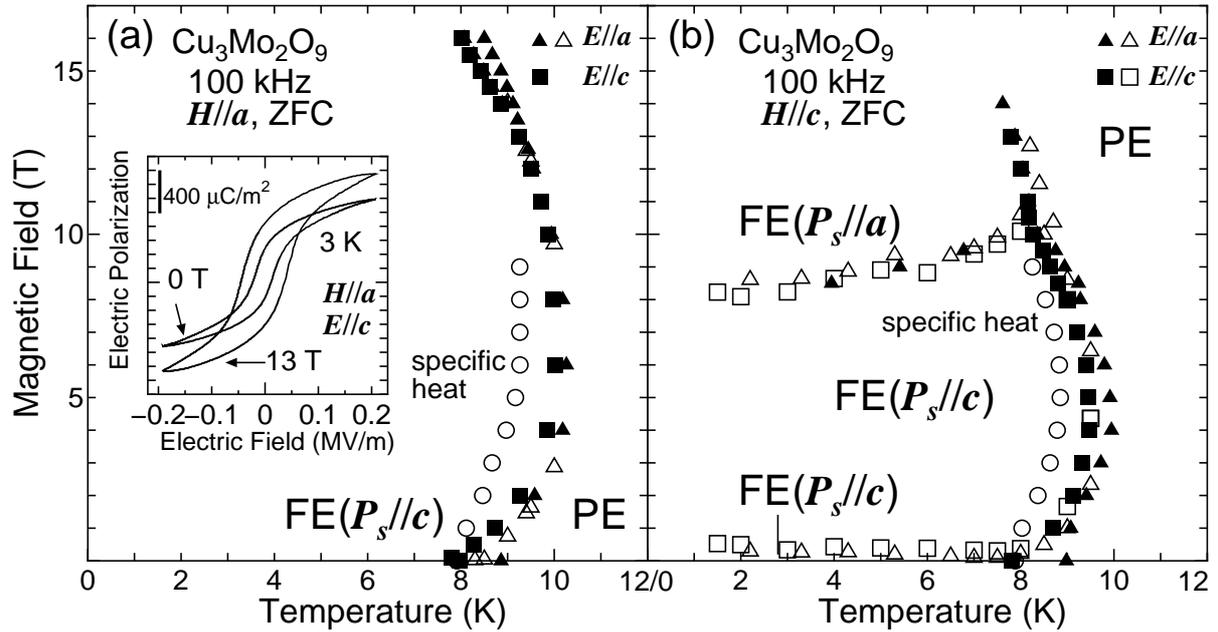}\hspace{2pc}%
\begin{minipage}[b]{38pc}
\caption{\label{H-Tphase}The $H-T$ phase diagrams in Cu$_3$Mo$_2$O$_9$. 
The shape of symbols distinguish the physical quantities 
to be used to obtain the phase boundary. 
The triangles, squares and circles denote the dielectric constants 
along the $a$ and $c$ axes and specific heat, respectively.
The solid (open) symbols denote the phase boundary
obtained from the data of the $T$ ($H$) dependence.
The inset of (a) shows typical polarization-electric field loops 
when the electric field is applied along the $c$ axis at 3 K under 0 and 13 T.
}
\end{minipage}
\end{figure}

\medskip
\section{Results and Discussion}

Figures 2(a) and (b) show the typical $T$ dependences
of $\epsilon_\alpha$ ($\alpha = a$ or $c$) 
under a fixed $H$ along the $a$ axis ($H_{a}$),
(the $\epsilon_\alpha$-$T$ curves),
respectively, each of which has a (local) maximum value
$\epsilon_\alpha^{\rm peak}$
at $T_\alpha^{\rm peak}$. 
$T^{\rm peak}_\alpha$ 
increases at 6 T, but it decreases at 16 T.
$\epsilon_a$ at 16 T shows two anomalies consisting of two peaks, as shown in Fig. 2(a).
As shown in Fig. 2(c), the detailed $\epsilon_a$-$T$ curves 
under magnetic fields between 12 and 16 T, 
the $\epsilon_a$ above 15 T has two peaks indicated by arrows.
The values of $T^{\rm peak}_\alpha$ 
against $H$ are plotted in the $H$-$T$ phase diagram 
in Fig. 3(a) by the solid symbols.
Figure 2(d) shows the typical $H$ dependences of $\epsilon_a$
from 8 to 12 K (the $\epsilon_a$-$H_a$ curves). 
The $\epsilon_a$-$H_a$ curves have 
a cusp between 8.3 and 9 K and two ones between 9.5 and 10 K, respectively. 
These are plotted in
the phase diagram in Fig. 3(a) by the open symbols.

The inset of Fig. 3(a) shows the 
polarization-electric field ($P_c$-$E_c$) loops at 3 K under 0 and 13 T 
when the electric field $E$ and the magnetic field $H$ are applied along the $c$ and $a$ axes, respectively.
Typical ferroelectric $P_c$-$E_c$ hysteresis loops were observed, indicating that 
the AFM phase is ferroelectric and has a spontaneous electric polarization 
parallel to the $c$ axis.
This result is consistent with the fact that the peak height of $\epsilon_c$
is about ten times larger than that of $\epsilon_a$, as seen in Figs. 2(a) and (b).
At 13 T, all of the saturated polarization, the spontaneous polarization and the coercive electric field are larger than the values at 0 T. 
These results indicate that the ferroelectric correlation increases with increasing $H$. 
These magnetic-field dependences of electric polarization are different from the ones when the magnetic field is applied along the $c$ axis. The magnetic-field dependence of magnetization also depends on the direction of the magnetic field. 
We consider that the anisotropic electric property of Cu$_3$Mo$_2$O$_9$ is related to the anisotropic magnetization of this compound.\cite{Hamasaki2008,Hamasaki2010} 
And it suggests the ferroelectricity originating from the frustrating spin configuration as an origin of the multiferroic behavior.

Together with the phase
boundary obtained from the $T$ dependence of the specific
heat under $H$, 
we obtain the ($H$-$T$) phase diagram
in Fig. 3(a) when $H$ is applied parallel to the $a$ axis.
The $T_{\rm N}$ obtained by the specific heat is little bit lower than 
$T_\alpha^{\rm peak}$.

We compared Fig. 3(a) to the $H$-$T$ phase diagram 
from ref. \cite{Kuroe2011-2} (Fig. 3(b)), 
which is obtained under $H$ along the $c$ axis.
There are three different ferroelectric phases 
in the AFM phase of Fig. 3(b),
indicating that the phase diagram for $H//a$ is simpler than that for $H//c$.
When $H//c$,
a change in direction of the spontaneous electric polarization
occurs at the phase boundary running from ($H, T$)
= (8 T, 2 K) to (10 T, 8 K).
Around the tricritical point at 10 T and 8 K, 
the change of the direction in the electric polarization 
causes a colossal magnetocapacitance effect.\cite{Kuroe2011-2}  
When $H//a$, the $\epsilon_{a}$-$T$ curve shows two peaks 
under magnetic fields above 14 T,
as shown in Fig. 2(c),
suggesting that a new phase appears in the narrow region.
Then, another tricritical point may exist around (9 K, 14 T).
At this tricritical point,
the strong magnetocpapacitance effect 
has not been observed in the present work.
This suggests that 
the strong magnetocapacitance effect 
originates from the change in the direction of the spontenous electric polarization.
We conclude that much different multiferroic behaviors occur
when $H$//$a$ and $H$//$c$.
At present we are interested in the multiferroic
behaviors under high magnetic fields.

\section*{Acknowledgments}
\ This work is partly supported by a Grants-in-Aid for
Scientific Research (C) (No. 40296885) and 
on Priority Area (No. 19052005) from the Ministry
of Education, Culture, Science and Technology of Japan
(MEXT).
We also wish to acknowledge the technical assistance
of Mr. R. Kino and Mr. M. Suzuki.

\smallskip


\section*{References}
\medskip
\begin{thebibliography}{99}
\bibitem{Kimura2003} Kimura T, Goto T, Shintani H, Ishizawa K, Arima T and Tokura Y 2003 {\it Nature} \textbf{426} 55
\bibitem{Cheong2007} For review, Cheong S W and Mostovoy M 2007 {\it Nature Mat.} \textbf{6} 13 
\bibitem{Katsura2005} Katsura H, Nagaosa N and Balatsky A V 2005 {\it Phys. Rev. Lett.} \textbf{95} 057205
\bibitem{Kenzelmann2005} Kenzelmann M, Harris A B, Jonas S, Broholm C, Schefer J, Kim S B, Zhang C L, Cheong S W, Vajk O P and Lynn J W 2005 {\it Phys. Rev. Lett.} \textbf{95} 087206
\bibitem{Mostovoy2006}  Mostovoy M 2006 {\it Phys. Rev. Lett.} \textbf{96} 067601
\bibitem{Arima2005}  Arima T, Goto T, Yamasaki Y, Miyasaka S, Ishii K, Tsubota M, Inami T, Murakami Y and Tokura Y 2005 {\it Phys. Rev.} B \textbf{72} 100102R
\bibitem{Kuroe2011-2} Kuroe H, Hosaka T, Hachiuma S, Sekine T, Hase M
, Oka K, Ito T, Eisaki H, Fujisawa M, Okubo S and Ohta H to be published in J. Phys. Soc. Jpn.
\bibitem{Hamasaki2008} Hamasaki T, Ide N, Kuroe H, Sekine T, Hase M, Tsukada I and Sakakibara T 2008 {\it Phys. Rev.} B \textbf{77} 134419
\bibitem{Hamasaki2010} Hamasaki T, Kuroe H, Sekine T, Akaki M, Kuwahara H. and Hase M 2010 J. Phys.: Conf. Ser. \textbf{200} 022013
\bibitem{Kuroe2010} Kuroe H, Hamasaki T, Sekine T, Hase M, Oka K, Ito T, Eisaki H and Matsuda M 2010 J. Phys.: Conf. Ser. \textbf{200} 022028.
\bibitem{Kuroe2011} Kuroe H, Hamasaki T, Sekine T, Hase M Oka K, Ito T, Eisaki H, Kaneko K, Metoki N, Matsuda M and Kakurai K 2011 {\it Phys. Rev.} B \textbf{83} 184423
\bibitem{Kimura2006} Kimura T, Lashley J C and Ramirez A P 2006 {\it Phys. Rev.} B \textbf{73} 220401R
\bibitem{Hase2008} Hase M, Kitazawa H, Ozawa K, Hamasaki T, Kuroe H and Sekine T 2008 {\it J. Phys. Soc. Jpn.} \textbf{77} 034706
\end{thebibliography}
\end{document}